\documentclass[12pt,preprint]{aastex}

\def\kms{\ifmmode{\rm km\thinspace s^{-1}}\else km\thinspace s$^{-1}$\fi}
\def\ms{\ifmmode{\rm m\thinspace s^{-1}}\else m\thinspace s$^{-1}$\fi}

\shorttitle{THE EFFECT OF BINARITY ON STELLAR ROTATION}
\shortauthors{Meibom et al.}

\begin{document}

\title{THE EFFECT OF BINARITY ON STELLAR ROTATION
- BEYOND THE REACH OF TIDES\altaffilmark{1}}

\author{S{\o}ren Meibom\altaffilmark{2,3,4},
Robert D. Mathieu\altaffilmark{4}, and Keivan G. Stassun\altaffilmark{5}}

\altaffiltext{1}{WIYN Open Cluster Study. XXXI.}
\altaffiltext{2}{{\it smeibom@cfa.harvard.edu}}
\altaffiltext{3}{Harvard-Smithsonian Center for Astrophysics,
60 Garden Street, Cambridge, MA - 02138}
\altaffiltext{4}{Astronomy Department, University of Wisconsin - Madison,
Madison, WI - 53706}
\altaffiltext{5}{Physics and Astronomy Department, Vanderbilt University,
Nashville, TN - 32735}


\begin{abstract} \label{abs}

We present a comparison between the rotation period distributions
of solar-type single stars and primary stars in close binaries
(0.1 AU $\la a \la$ 5 AU) in the young ($\sim$ 150 Myr) open
cluster M35 (NGC\,2168). We find that the primary stars in the
close binaries rotate faster than the single stars, on average.
The differences in the means and medians between the period
distributions are statistically significant at the 99.9\% level
or higher. The faster rotation among the primary stars in close
binaries is {\it not} due to tidal synchronization as tidally
evolved stars are excluded from the comparison. We discuss this
result in the context of different early-evolution accretion
processes and star-disk interactions for single stars and stars
in close binaries.

\end{abstract}



\section{INTRODUCTION}	\label{intro}

The rotational properties of young low-mass $(\la 2~M_{\odot})$
stars challenge our understanding of their angular momentum evolution,
and specifically of the physical mechanisms, internal and external,
that control transport of angular momentum. The importance of determining
these processes is underscored by the observed loss of angular momentum
during the pre-main-sequence (PMS) phase producing slowly rotating
zero-age main-sequence (ZAMS) stars, and by observations of orders
of magnitude dispersions and distinct mass dependencies in the rotation
period distributions for coeval samples of PMS, ZAMS, and main-sequence
(MS) stars \citep[e.g.][and references therein]{barnes03a,hm05}.

Inspired by these challenges a large body of observational
and theoretical work on the rotational evolution of low-mass
PMS and MS stars was accomplished over past decades. The
interpretations and discussions of the results have focused
primarily on the roles of stellar magnetic winds and interactions
between global stellar magnetic fields and circumstellar (CS)
disks in controlling the angular momentum evolution.

In the context of single stars, current models of PMS rotational
evolution \citep[e.g.][]{tpt02,bfa97,bs96} rely on the regulation
of stellar angular momentum by CS disks via magnetic disk-braking
\citep[e.g.][]{konigl91,sno+94} to recreate the observed rotational
properties of ZAMS stars from those of PMS stars. It is generally
assumed that the action of magnetic star-disk coupling is to force
the star to rotate at constant angular velocity for the lifetime
of the disk. It follows from these models that stars with massive
long-lived disks will reach the ZAMS rotating slower than stars
with short-lived disks.

In parallel, models of the evolution of CS disks in PMS binary
stars have studied the effect of a close companion on the sizes,
masses, and accretion rates, and thereby the lifetimes, of the disks.
In binaries with separations $\la 100$ AU, current models of
star-disk interactions predict that a companion star will truncate
\citep{ac96,al94,lp93}, cause accelerated mass-accretion from
\citep{pt95,kp95}, and potentially disrupt \citep{pringle91,artymowicz92}
the CS disks. Accordingly, most models predict that the lifetimes
of CS disks are reduced for stars in close binaries as compared
to stars in wider binaries and single stars (but see \citet{ac96}).
Such models find support in observational evidence for truncated
CS disks and ``gaps'' in in disks cleared by a companion
\citep[e.g.][and references therein]{jmf96,jm97}.

Together, the models suggest that stellar companions may affect
one another's rotational evolution indirectly, by virtue of their
disruptive effects on the CS disks that would otherwise act to
regulate stellar rotation. To test for a relationship between
binarity and rotation, an observational comparison of the rotational
evolution between young single and binary primary stars of the
same age is needed.

In the closest binary stars, tidal theory \citep[e.g.][]{zahn77,hut81}
predicts that stellar rotation is affected by tidal interactions. The
timescale for tidal influence on rotation is dominated by the binary
separation ($t_{sync.} \propto (a/R)^6$). Thus tidal theory predicts
a (time-dependent) binary separation beyond which rotational evolution
due to tides is negligible. Accordingly, in binaries with greater
separations, any effect of binarity on stellar rotation must originate
from processes other than tidal interactions between the two stars.

Previous searches for a relationship between binarity and rotation
among late-type dwarfs were published by \citet{brn97} and \citet{pgr+02}.
Both groups used projected rotational velocities ($v\sin i$) and
speckle imaging to identify binaries as close pairs (separations
$\sim$ 10-1000 AU). \citeauthor{brn97} found no evidence for a
relationship between binarity and rotation in the Pleiades.
\citeauthor{pgr+02} reported higher $v\sin i$'s of 4 binaries with
separations between 10-60 AU than for 12 wider binaries in $\alpha$
Persei.

We present an analysis based on coeval samples of spectroscopic
single stars and spectroscopic binaries, 83\% of which have
determined spectroscopic orbits \citep{mm05}. All are members
of the 150 Myr open cluster M35, and all single/primary stars
have well determined rotation periods \citep{mms06,meibom05}.
The sample of spectroscopic binaries probe a previously unexplored
domain of binary separations ($\sim$0.05-5 AU) in which the CS
disks are expected to be truncated, disrupted, and accreted onto
the stars on short timescales compared to CS disks of the single
stars or stars in wider binaries. Our samples are therefore well
suited to search for a relationship between binarity and rotation
on the MS, which is the goal of this study.

Section~\ref{obs} outlines the spectroscopic and photometric observations
and Section~\ref{samples} describes the identification of single and
close binary stars and the exclusion of tidally synchronized binaries.
We present and compare the distributions of rotation periods for single
and close binary stars in Section~\ref{distr} and summarize and discuss
our results in Section~\ref{summ}.


\section{OBSERVATIONS} \label{obs}

We have conducted two parallel observational programs of
late-type stars in the open cluster M35: 1) High precision
\citep[$\sim 0.5\,km/s$][]{mbd+01,gm07} radial-velocity surveys to
identify single and binary cluster members and determine orbital
parameters for the closest binaries; 2) Comprehensive photometric
time-series surveys to determine stellar rotation periods from
light modulation by star-spots.

All spectroscopic data were obtained over a one-degree field centered
on M35 using the WIYN 3.5m telescope\footnote{http://www.noao.edu/wiyn}
with the Hydra Multi-Object Spectrograph. Approximately 125 spectroscopic
binary members have been identified from the radial-velocity survey
including 32 with orbital periods between 2.25 to 3112 days ($0.04
\la a \la 5 AU$, assuming $M_{p} = 1~M_{\odot}$ and $M_{s} = 0.75~M_{\odot}$;
\citet{mm05}). Radial-velocity membership probabilities were calculated
based on the formalism by \citet{vkp58} and Gaussian fits to the distinct
peaks in the radial-velocity distributions of the cluster ($-8.1\kms$)
and field stars \citep[see][]{mms06,mms07b}.

The photometric data were obtained within the one-degree field of
the spectroscopic survey using the WIYN 0.9m telescope\footnote{
http://www.noao.edu/0.9m}. Photometric members were selected based
on their location in the color-magnitude diagram (CMD) within or
above the main sequence (allowing for inclusion of equal-mass binaries).
A relative photometric precision of $\sim$ 0.5\% was obtained for
the brightest stars ($12 \la V \la 15$). We have determined stellar
rotation periods for 196 photometric and radial-velocity members
of M35 \citep{meibom05,mms07b}. The periodic variability detected
in the light curves of single-lined spectroscopic binaries is caused
by spots on the primary stars, and is therefore a reliable measure
of their rotation periods \citep[see][for a detailed discussion]{mms06}.


\section{DEFINING BINARY AND SINGLE STAR SAMPLES} \label{samples}

Of the 196 photometric and spectroscopic members of M35 with rotation
periods, 118 have three or more radial-velocity measurements,
allowing us to test for variability indicative of a close companion
star. Specifically, we can distinguish between binaries with $a \la 5$
AU and single stars or primary stars in wider binaries (both of which
are hereinafter called single). 

We apply the following criteria for determining whether a star
is single or a member of a close binary system:

\indent Single star: $N_{RV} \ge 3$, $\sigma_{RV} \le 0.5~km~s^{-1}$,
and photometrically single.

\indent Binary star: $N_{RV} \ge 3$ and $\sigma_{RV} \ge 1.5~km~s^{-1}$.

\noindent $N_{RV}$ denotes the number of radial-velocity measurements, 
$\sigma_{RV}$ is the standard deviation of these measurements, and 
``photometrically single'' refers to a location on the ``single''
star cluster sequence in the M35 CMD, thereby excluding equal brightness
binaries.

These criteria ensure that single stars are not variable above the 1
$\sigma$ level, and binary stars vary above the 3 $\sigma$ level
in radial velocity. Stars in the grey-zone between single and
binary ($0.5~km~s^{-1} \ga \sigma_{RV} \la 1.5~km~s^{-1}$) were
not considered. The radial-velocity thresholds are independent
of stellar rotation as we find no correlation between stellar
rotation period and $\sigma_{RV}$ for either the single stars
or stars in the grey-zone. 

We find 53 single stars and 18 binary stars with measured rotation
periods. Of the 18 binary stars, 15 have determined spectroscopic
orbits \citep{mm05}. Of the 3 remaining binaries without spectroscopic
orbits, 2 have $\sigma_{RV}$'s of $8.3~km~s^{-1}$ and $5.9~km~s^{-1}$,
based on $N_{RV} = $ 24 and 12, respectively, significantly above
the $1.5~km~s^{-1}$ threshold. 
The lack of satisfactory orbital solutions for these two binaries
is due in part to blending of the primary and secondary components,
preventing accurate determination of the velocities, and to insufficient
sampling of the orbits near periastron passage in moderate to high
eccentricity orbits. Preliminary orbits have been found for both,
but additional measurements are needed to confirm and constrain
the solutions.
The last binary without an orbit has a $\sigma_{RV}$ of $1.6~km~s^{-1}$
based on 4 velocities. Inspection of each individual spectrum and
cross-correlation function leaves no reason to doubt the quality of
the derived radial velocities for this star and we keep it in the
binary sample.

By design, our spectroscopic observing program targets the closest
binary stars and gives lowest priority to stars with no or low-amplitude
velocity variations. Accordingly, the 53 stars identified as single
have only between 3 and 6 velocity measurements separated in time
by $\sim$0.5 to 1.5 years. We cannot rule out the possibility of
long-period low brightness ratio binaries among the single stars.
However, we find from Monte Carlo analysis based on 30,000 binary
orbits, assuming an overall binary fraction of two thirds for
solar-type stars \citep[e.g.][]{mrd+92,dm91}, that of the 53 stars
in the single sample, none will be binaries with semi-major axes
less than 10 AU, and at most 7 (13\%) will be binaries with separations
less than 100 AU. We sampled each binary orbit with a frequency
similar to our actual observing pattern.

{\it Excluding tidally evolved binary stars}:
Because we are interested in effects on stellar rotation in
close binaries other than those imposed by tidal interactions,
we evaluate the degree of tidal evolution in the closest binaries
and exclude from our sample those likely affected by tidal
synchronization at the age of M35 ($\sim$ 150 Myr).

Five binaries have orbital periods similar to or shortward of
the tidal circularization period for M35 \citep[$P_{circ} = 10.1$
days, $a \simeq 0.12$ AU;][]{mm05}. Considering theoretical predictions
that the rate of tidal synchronization exceeds that of tidal circularization
by a factor $\sim 10^{2}$ for constant stellar interior structure
\citep{zahn89,hut81}, tidally synchronized or pseudo-synchronized
stellar spins are expected for all 5 primary stars. However, time-series
spectroscopic and photometric observations of the 5 binaries reveal
that only two have synchronized primary stars and circular orbits
\citep{mms06}. The primary stars in the remaining three binaries,
only one of which has circularized, are rotating either highly super-
or sub-synchronous. 

Theoretical tidal synchronization times for the 5 binaries can be
estimated from the prescription by \citep{hut81}. The synchronization
times for the 5 binaries range from $\la$ 10 Myr to several Gyr
due to their differences in orbital period and eccentricity, and
in the mass of the primary star. At this time there is no consistent
agreement between the expectations and predictions of tidal theory
and the observed levels of tidal synchronization and circularization
in the 5 closest binaries in M35. Therefore, in the analysis that
follows we will exclude all 5. In the resulting sample of 13 binaries
the shortest orbital period is 30.13 days, and the influence of tidal
interactions at the age of M35 can safely be ignored.


\section{THE SINGLE- AND BINARY-STAR ROTATION PERIOD DISTRIBUTIONS}
\label{distr}

\begin{figure}[ht!]
\epsscale{1.0}
\plotone{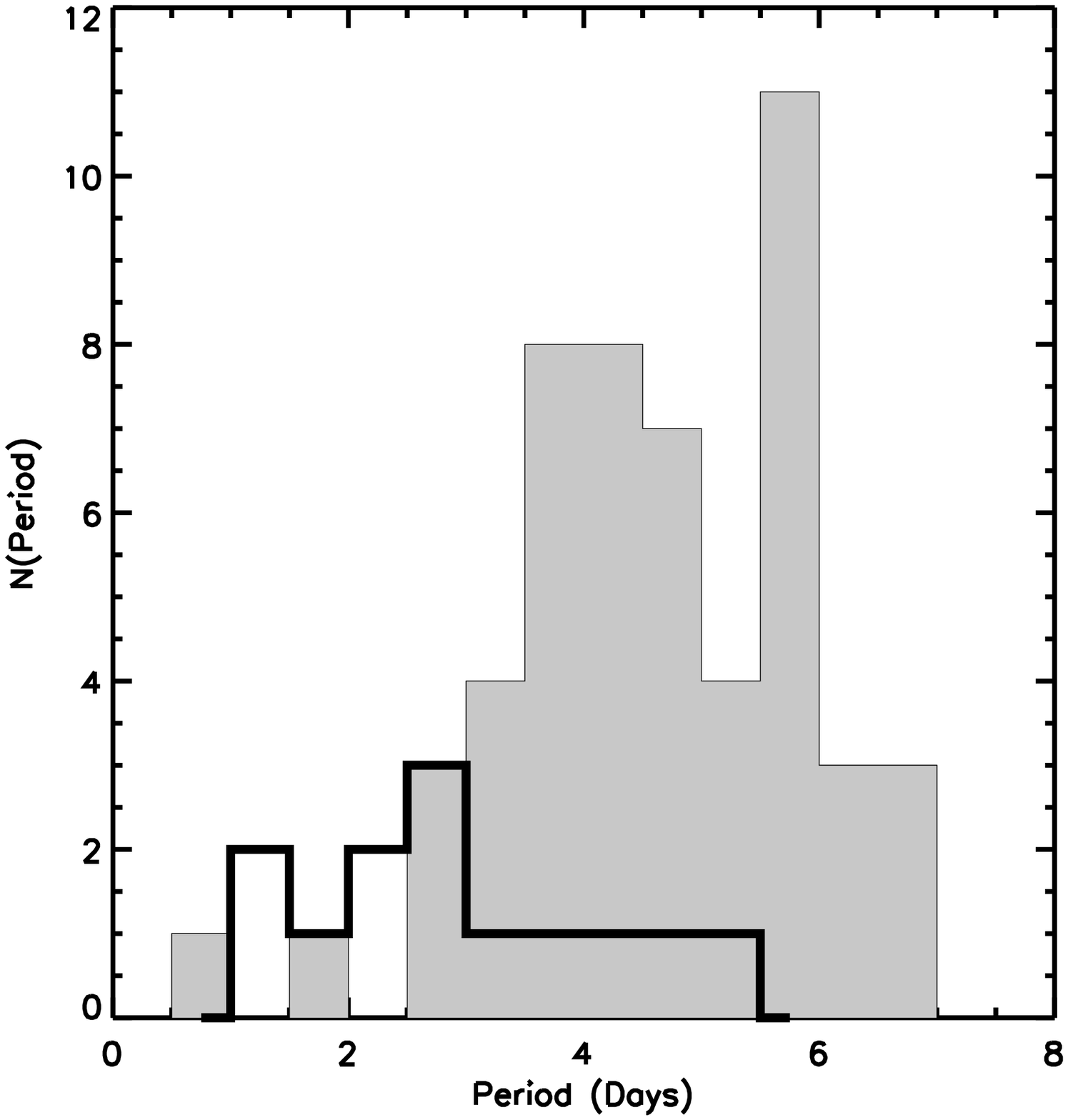}
\caption{Rotation period distributions for 53 single stars (grey histogram;
$\overline{P_{rot}} = 4.64$ days, $\tilde{P_{rot}} = 4.70$ days) and
13 binary primary stars excluding the five shortest period binaries (solid;
$\overline{P_{rot}} = 2.95$ days, $\tilde{P_{rot}} = 2.83$ days).
\label{svph}}
\end{figure}

Figure~\ref{svph} shows the rotation period distribution of the single
star sample (grey histogram) and for the sample of binary primary stars
(solid line histogram). The mean and median rotation periods of the sample
of binary primary stars fall 1.7 days and 1.9 days, respectively, short
of the mean and
median of the single star sample. The significance of the differences
in the mean and median rotation periods of the single star and the
binary star samples can be formally evaluated by the ``Student's''
t-test and the Mann-Whitney u-test \citep{ptv+92}. These two statistical
tests are parametric and non-parametric tests of the null hypothesis
that two populations derive from the same parent population, or
equivalently that the differences in the means or medians between
two distributions are not statistically significant. We performed
both tests on the single star period distribution against the binary
primary star period distribution. Both tests result in less than 0.1\%
probability that the difference in the means/medians is by chance and
thus that the two distributions derive from the same parent distribution.


\section{SUMMARY AND DISCUSSION}	\label{summ}

We compare the rotation period distributions of solar-type single
stars and primary stars in close ($a \la 5$ AU) binaries in the
150 Myr open cluster M35. We find that the primary stars rotate
faster than the single stars or primary stars in wider binaries,
on average. This relationship between binarity and rotation is not
due to tidal synchronization, and we find no correlations between
stellar rotation and orbital period or eccentricity among the close
binaries.

The observed effect of on average faster rotation in close
binaries is consistent with a model scenario involving truncation
of the CS disk lifetime by a close companion and consequently
a shortened phase of magnetic disk-braking of the stellar rotation
during the early PMS phase. Whether magnetic disk-braking is the
dominant process setting stellar rotation at 150 Myr remains
uncertain on both observational and theoretical grounds.

Conceivably, the observed difference in rotation at 150 Myr
may reflect differences between single and close binary stars
in the amount and distribution of angular momentum at their
formation and very early evolution. The formation of the closest
binaries, in particular, may differ significantly from the
formation of single stars or stars in wide binaries
\citep[][and references therein]{larson03}.
For example, highly variable accretion rates are frequently
associated with the presence of a close companion, and
observations of jet-like outflows (Herbig-Haro and FU Orionis
stars) have been linked to episodic accretion caused by
tidal interactions between proto-stars and their disks in
close binaries \citep{reipurth01,hk96,bb92}. A radically
different circumstellar environment between single stars
and stars in close binaries is carried through to the PMS
phase where single stars accrete from extensive CS disks
while stars in close binaries may accrete through different
processes such as accretion streams \citep{al96,msb+97,jds+07}.
Whether these different accretions processes also lead to
different depositions of angular momentum remains to be seen.

If indeed the observed rotational difference between single and
primary stars in M35 derive from differences in the proto-stellar
environment or from star-disk interactions in the early PMS
phase, it should be more pronounced in younger stellar populations
and gradually disappear for older populations as magnetic winds
spin down all stars not tidally locked. If on the other hand the
rotational difference is persistent over stellar age, then it may
be caused by an as-yet-unknown effect of a close companion.
Further observational study of the single- and binary-star
rotation periods in younger and older populations will probe
any dependency on stellar age.


\acknowledgments

We thank UW-Madison and NOAO for time granted on the WIYN
telescopes, and recognize the exceptional and friendly support
from staff and observers at both sites. We thank the referee
for suggestions that strengthened the manuscript and the reported
result. This work was supported by NSF grant AST 97-31302,
and by a Cottrell Scholarship from the Research Corporation
to K.G.S.




\clearpage

\end{document}